\documentclass[a4paper]{llncs}
\usepackage[utf8]{inputenc}
\usepackage[english]{babel}
\usepackage{hyphenat}
\usepackage{amsmath}
\usepackage[pdftex]{graphicx}
\usepackage{tabularx}
\usepackage{calc}
\usepackage{pslatex}
\usepackage{listings}
\usepackage{url}
\usepackage{caption}
\usepackage{subcaption}
\usepackage{footnote}
\usepackage{color}
\usepackage{todonotes}

\usepackage{xcolor}
\usepackage{colortbl}
\usepackage{booktabs}
\usepackage{tabu}

\usepackage{tikz}
\usepackage{siunitx}

\usepackage{paralist}

\newcommand{\ie}{\textit{i.e.},\ }
\newcommand{\eg}{\textit{e.g.},\ }

\newcommand{\mapenc}{\textit{Map/Encap}}

\newcommand{\eipi}{EIP}
\newcommand{\rgp}{RGP}

\begin{document}

%\title{Preventing DDoS with Cryptographically Generated Ephemeral IP Identifiers}
\title{EIP \--- Preventing DDoS with Ephemeral IP Identifiers Cryptographically Generated }

%\author{
%Ricardo Paula Martins, Jos\'{e} Legatheaux Martins, Henrique Jo\~{a}o Domingos
%
%NOVA LINCS, Department of Computer Science, Faculty of Science and Technology,\\
%Universidade NOVA de Lisboa, 2829--516 Caparica, Portugal\\
%Email: rmp.martins@campus.fct.unl.pt, jose.legatheaux@fct.unl.pt, hj@fct.unl.pt
% }

\author{Ricardo Paula Martins, Jos\'{e} Legatheaux Martins, Henrique Jo\~{a}o Domingos}
\tocauthor{}

\institute{NOVA LINCS, Department of Computer Science, Faculty of Science and Technology,\\
Universidade NOVA de Lisboa, 2829--516 Caparica, Portugal\\
Email: rmp.martins@campus.fct.unl.pt, jose.legatheaux@fct.unl.pt, hj@fct.unl.pt
 }

\maketitle

\begin{abstract}

Nowadays, denial of service (DoS) attacks represent a significant fraction of all
attacks that take place in the Internet and their intensity is
always growing. The main DoS attack methods consist of flooding their victims
with bogus packets, queries or replies, so as to prevent them from fulfilling
their roles.
Preventing DoS attacks at network level would be simpler if
% making use of
end-to-end strong authentication in any packet exchange
was mandatory.
However, it is also
likely that its mandatory adoption would introduce more harm than benefits.

In this paper we present an end-point addressing scheme and a set
of security procedures which  satisfy most of network level DoS prevention requirements.
Instead of being known by public
stable IP addresses, hosts use ephemeral IP Identifiers cryptographically generated
and bound to its usage context. Self-signed certificates and
challenge-based protocols allow, without the need of any third
parties, the implementation of defenses against DoS attacks.
Communication in the open Internet while using these special IP addresses is
supported by the so-called {\mapenc} approaches, which in our point of
view will be sooner or later required for the future Internet.

\end{abstract}

%\begin{IEEEkeywords}
%Denial of Service attack, DoS, DDoS, Map-Encap, network security, client puzzle
%\end{IEEEkeywords}

\section{Introduction}
\label{introd}

Nowadays, denial of service (DoS) attacks represent a significant fraction of all
attacks that take place in the Internet~\cite{akamai-q3-2015,verisign2014}. Their main goal
is to preclude sites and network infrastructure  subsystems from
providing service, for economical, political or vandalism goals.
DoS as a service can now be purchased from the Internet
underground by some tens of dollars~\cite{booters} for the smallest attacks.
The main DoS attack methods consist of flooding their victims
with bogus packets, queries or replies, so as to prevent them of fulfilling
their roles. Floods with tens of Gbps are quite common~\cite{akamai-q3-2015}.

Many attacks are performed at the application level, flooding sites with
otherwise regular service queries. However, those requiring less
attack power are infrastructure level attacks, based on stateless, unauthenticated query/reply
protocols, like ICMP, SNMP, NTP, DNS or SSDP~\cite{ampot}.
Most of the deadliest ones use a set of infected machines under the
control of the attacker, known
as a botnet, to perform what is called a Distributed Denial of Service (DDoS)
attack~\cite{survey-ddos1}. If the botnet sends IP packets with the victim's address as source
 address to intermediate servers, known as reflectors, the latter will flood the victim with
 service replies, effectively hiding the botnet participants. This
is called a~\emph{Reflection Attack} (DRDoS). When the reply has many more bytes than the
request, the attack may become quite deadly for a relative cheap investment,
and is known as an~\emph{Amplification Attack}.

The main characteristics of the current Internet ecosystem being explored to
perform DoS attacks are:
\begin{inparaenum}
\item the absence of mandatory IP
address authentication and user authentication;
% \item the absence of detailed resource accounting and the absence
% of the need of previous authorization to send packets to any reachable receiver;
\item the absence of both detailed resource accounting and mandatory previous authorization to send packets to any reachable receiver;
\item system software bugs; and finally,
\item many service definitions were introduced when all internet
users were ``good net citizens'', lacking basic security precautions,
and allowing their leveraged to grow botnets and reflection attacks.
\end{inparaenum}
On the other hand, by lowering barriers against new users
and providers, these weakness play a pivotal role
in the never ending expansion of the Internet and in lowering
its operating costs.

%%%%%%%%%%%%%%%%%%%%%%%%%%%%%%%%%%%%%%%%%%%%%%%%%%%%%%%
%\commenti{Esta frase podia ser dispensável se não tivesse uma importância tão grande para o resto do artigo -- conclusões?}
%%%%%%%%%%%%%%%%%%%%%%%%%%%%%%%%%%%%%%%%%%%%%%%%%%%%%%%

Simple countermeasures against many types of DoS attacks would be
possible if providers prevented their customers from using fake
source IP addresses, or implemented context specific analysis
of their actions and raised barriers against unusual activities (\eg reverse firewalls).
However, providers have no incentive to perform these actions
and some of them could be against the law and the users' freedom.

Implementing countermeasures in the core of Internet or
requiring a close cooperation of most providers is also neither
scalable nor very realistic. Thus, most countermeasures are
nowadays implemented near the potential victims, where there is
a clear incentive. Current practices
include server replication and bandwidth increase, as well
as resorting to security providers that use expensive
and attack-vector dependent detection techniques, in a never ending
``cops and robbers'' chase ~\cite{survey-ddos1}.

Even when DoS attacks do not prevent services from being offered,
they clearly increase the resources needed to maintain the
quality of service thus indirectly increasing their cost.
Additionally, due to the absence of effective measures against these
attacks, network administrators tend to be very conservative, suspect
all traffic but the most trivial one, and block all new protocols. This
state of affairs also prevents innovation, blocks the end-to-end
nature of the network and deeply contributes to the so called
ossification of the Internet~\cite{rethinking}.

%Preventing DoS attacks would be simpler if TCP/IP networks were connection oriented,
%(\eg like circuit switching) and all users were authenticated before
%being allowed to send any packet. However, the former is in fact not desirable
%at all since the stateless nature of the Internet is of paramount importance
%for its scalability. In what concerns user authentication, IPSec could be mandatory used.
%However, although being standardized since more then 15 years ago, it never took off
%outside the enterprise level. Its generalized adoption by parties previously unknown to one another
%requires the universal adoption of third-party issued certificates.
%Not only this seems currently not realistic since it has never been widely adopted, but it is also
%likely that its mandatory adoption would introduce, in our opinion, more harm than benefits.
%Security requires diversity of mechanisms and policies as well
%as close adaptation to the context. In fact, most infrastructure DoS attacks could be more
%easily defended if it was possible to:

Preventing DoS attacks would be simpler if all services
could only be implemented on top of secured and authenticated connections.
However, this simplistic approach would
negatively impact the performance of all infrastructure
services (\eg DNS, NTP, \dots) and popular TCP-based services,
and would introduce, in our opinion, more harm than benefits
related to third party certification management.
Security requires diversity of mechanisms and policies,
as well a balanced relation among risks, increased costs and
incentives for adoption.
In fact, most infrastructure DoS attacks could be more
easily defended if it was possible to:

\begin{itemize}
\item check if the source addresses of IP packets seems consistent or
at least sensible;
\item easily check with no false negatives
nor false positives if a packet is part of an attack;
\item guarantee that some endpoint is at the network location it pretends to be;
\item guarantee that transport and application level attacks are more expensive
to the attacker than to the attacked part; and
\item prevent parties that do not provide services to the open
public from being forced to use long term, stable, public addresses,
belonging to an IP subnet that can be scanned.
\end{itemize}
%
%These requirements are orthogonal to real end-to-end strong authentication,
%which is a more challenging requirement, related only to the semantics
%and context of the service at hand,
%not to DoS defense requirements.

In this paper we present an endpoint addressing scheme and a set
of security procedures we call  {\eipi} (Ephemeral IP identifiers cryptographically generated)
from now on, which satisfies most of the above goals.
Instead of being known by stable public
IP addresses, hosts use ephemeral IP identifiers cryptographically generated
and bound to their usage context. Self-signed certificates and
challenge-based protocols allow the implementation of the above defense mechanisms
without the need of any third parties.
Communication in the open Internet while using these special IP addresses is
supported by the so-called {\mapenc} approach \cite{lisp}, which we will present in the
next section.

Today, ``DDoS as a service'' is provided for ridiculously cheap
prices when implemented by reflection attacks leveraging connectionless services, or
TCP SYN Flood attacks. This proposal is specially targeted at making these attacks very
expensive, making them impractical for the average attacker. 
Thus, the typical attacker is an individual or an organization capable of,
or using a botnet capable of, sending spoofed source IP packets. Nevertheless,
the proposal also helps the combat against transport and application level attacks.

We especially
do not address the scenario where very powerful organizations, have enough
resources to access the source addresses of arbitrary packets crossing
any network. The proposed addressing scheme makes address guessing
practically impossible.
The proposal is a mitigation mechanism, not an inexpugnable wall,
that tries to balance risks, costs and incentives. 

In Section~\ref{proposal} we present the cryptographic
mechanisms and protocols we propose, and in Section~\ref{evaluation}
we assess their effectiveness.
Next, in Section~\ref{costs}, we discuss the computational costs of
the selected cryptographic algorithms, and show that they are suitable
for the current generation of Internet hosts.
%In this section we
%also analyze how {\eipi} increases the monetary cost of
%attacks thus restricting their feasibility.
Section~\ref{related}
discusses the related work. The paper ends with the final conclusions
and future work, Section \ref{conclusions}.

\section{Locator / Identifier Separation}
\label{mapencap}

The locator / identifier split solutions, also called {\mapenc} solutions,
were motivated by the routing problems brought by mobility
and by scalability concerns, \eg~\cite{hip,lisp}.
They are characterized by the introduction of
two different name spaces: an identifier namespace, with the same
format as IP addresses, independent of the
communication end-points locations; and a locator namespace, \ie
the current IP address space, related to network location.
Communication between parties identified by identifiers
 takes place using tunnels ending at locators, by encapsulating
 IP packets sent between identifiers in packets sent between
 locators. This allows the application and transport layers to communicate
by the way of identifiers, free from specific constraints of network particularities,
like location, network service providers' address space or BGP routing issues.

Hosts can change locations, interfaces and tunnels, without changing their
identifiers and without disrupting ongoing sessions.
Thus, mobility, traffic steering using multiple communication endpoints, and multi-homing can be
dealt with in innovative new ways. In the rest of the section we will use
LISP (Locator / Identifier Separation Protocol)~\cite{lisp}
as an illustration of a {\mapenc} approach.

%\subsection*{Locator / Identifier Separation Protocol}

\begin{figure}[h]
\centering
\includegraphics[scale=0.35]{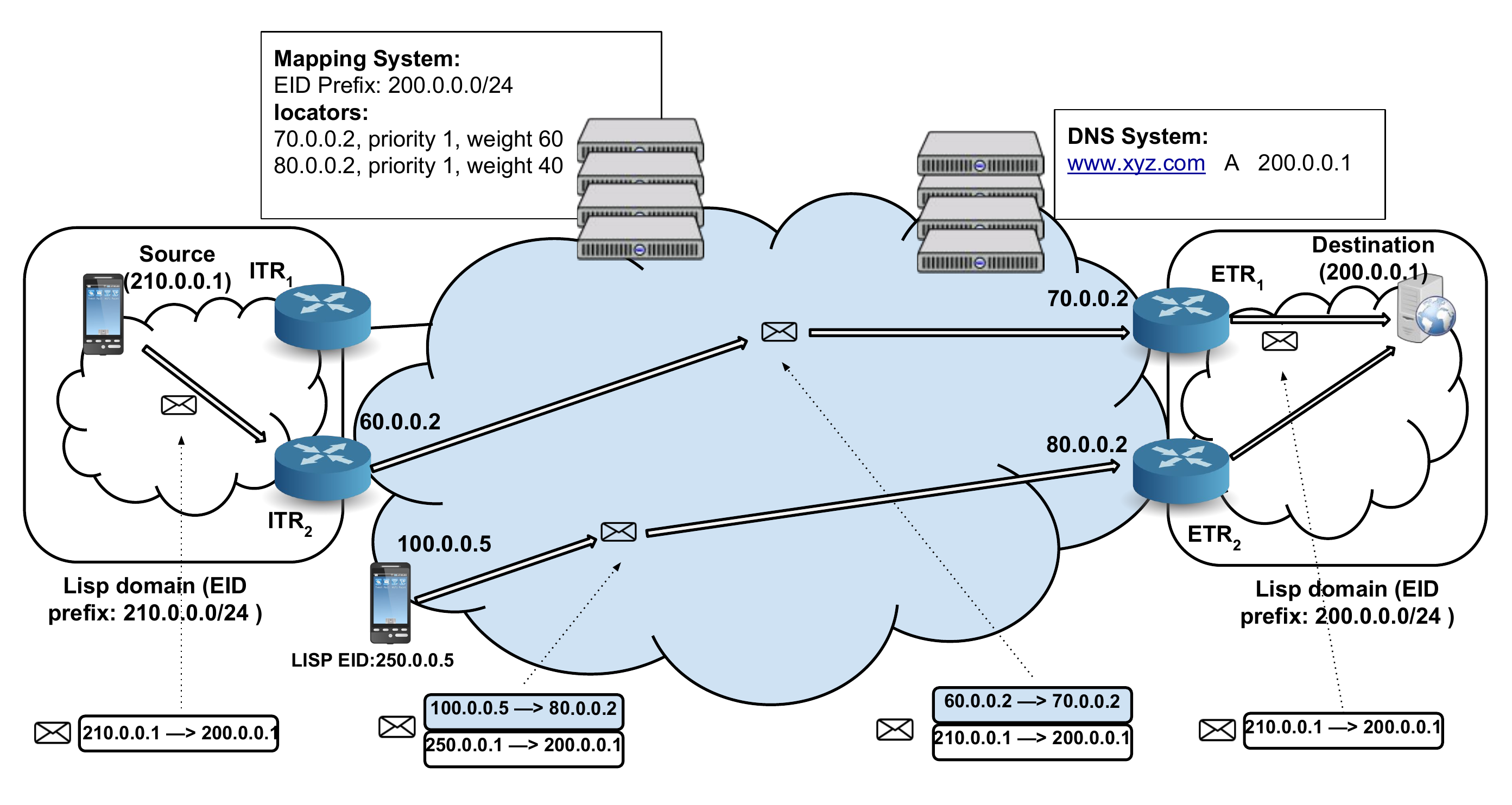}
\caption{The LISP {\mapenc} solution (ITR - Ingress Tunnel Router,
ETR - Egress Tunnel Router)}
\label{fig-lisp}
\end{figure}

In a LISP-based Internet, 
each customer network (\eg an AS or Autonomous
System)  is assigned an Endpoint Identifier (EID) prefix.
These EIDs are only routable inside each customers' network.
However, in order for an AS to send packets to another AS,
the packet has to cross the Default Free Zone, where EIDs are not routable.
For this purpose, the border routers on the AS encapsulate packets leaving that
AS with the Routing Locator (RLOC) of the tunnel end.
These encapsulated packets will traverse the Internet until they reach a
border router of the intended AS, where the external header will be removed.
Since the identifier in the inner header is routable in this AS,
the original packet is delivered to its destined host (see Figure \ref{fig-lisp}).

The LISP infrastructure is composed of tunnel routers and the mapping  system.
The last implement the EID-to-RLOC mappings, required by the former,
using an hierarchy of mapping servers close to the DNS one.

%The LISP infrastructure is composed by the following set of devices:
%(1) Ingress Tunnel Router: placed at the border of an AS,
%which obtains EID-to-RLOC mappings by questioning Map Servers,
%and then encapsulates outbound packets with the RLOCs received.
%(2) Egress Tunnel Router: placed at the border of an AS,
%which receives inbound LISP packets, removes the outer header, and
%routes them to the AS network where the destination
%EIDs are routable.
%(3) Map Server: these servers implement the  EID-to-RLOC mappings.
%Different solutions have been used to implement this service.
%More recently, an hierarchy of map servers, closer to the DNS one, has been adopted.

% \commenti{Transformar este parágrafo focalizando-o no mapping system, porque é o único aspecto que interessa}

%The full LISP proposal encompasses another auxiliary map server (Map Resolver)
%and two interworking servers (Proxy ITR / ETR)  \cite{rfc6830,rfc6834}.

The above brief presentation follows the initial main motivation of the
LISP proposal: to address the scalability issues of the Internet core. However,
nothing prevents an host from acting like the end of the tunnels for its own identifiers,
and thus implementing an end-to-end {\mapenc} solution. That has been
adopted in the so called LISP Mobile Node proposal \cite{lisp-mn1}.
A LISP mobile node is a device using a lightweight implementation of a LISP domain
and acting as a tunnel's end for its EIDs. The node is assigned EIDs,
which never changes when it changes networks. However, the locators (RLOCs) attributed
are the ones provided by the device's local network interfaces. The proposal
supports fully end-to-end device and even network (EID prefix) mobility when the RLOCs are public
IP addresses. Support for NAT traversal has also been provided.

LISP adopted solutions for the issues brought by the proposal, which have
been sorted out using test beds
\cite{lisp} in close cooperation with major vendors. It is
now available in routers of some of
these vendors and normalized in several RFCs (\eg see RFC6830\--34)\footnote{
For lack of space, RFCs are only referenced along the paper by their numbers, since
they are easily accessible from the IETF.}.

The protocol is flexibly adaptable to different use cases.
For example, individual LISP EIDs can be IPv6 addresses,
self-generated, conforming to ORCHID (RFC 7343). In that case the mapping
system can be bypassed if the receiver of an encapsulated packet
trusts the sender EID / RLOC association, and answers directly
to the other extreme of the tunnel.
Also, nothing prevents a host from generating a new EID for each
new transport session it needs.
These kind of alternate uses are the cornerstones of our proposal,
as will be introduced in the next sections.

\section{IP addresses integrity based on self-certified ephemeral EIDs} %%%%%%%  identifiers?}
\label{proposal}
% Very brief overview of what the protocol consists of, goals

The {\eipi} proposal  introduces forge-resistant identifiers
exhibiting integrity proofs in IP packets, in order to
allow the victim to easily discern packets from DRDoS attacks. Those
identifiers are complemented with an optional client puzzle mechanism that confirms the
sender location before processing its message.

% Description of the basic protocol (headers and verification)

The proposal is based on ephemeral self-certified identifiers, valid for a few
minutes (\eg tens). It involves an additional security header added to the regular
LISP header containing metadata binding the identifier to its
verifiable current usage and location.
Thus, this proposal can be
incrementally adopted, and requires no changes to either applications or
transport protocols.

The goal is forcing the endpoint that initiates a transaction (e.g. an NTP client)
with another endpoint (e.g. an NTP server) to generate a specific source identifier,
$ID_{src}$, for that particular transaction. These sender identifiers are
IPv6 addresses conforming to ORCHID,
%These sender identifiers have the same
%format as IPv6 addresses, conforms to ORCHID
and are generated through cryptographic techniques.
%The receiver only accepts the first packet if it contains at least a valid certificate.

The receiver's identifier, $ID_{dst}$ (e.g. of the NTP server)
should be public (e.g. registered in
DNS) and registered in the mapping system, so that clients are able to
obtain one of its locators, $LOC_{dst}$, in a trustable way.
This registry confers integrity to the identifier $ID_{dst}$, since LISP's
mapping system requires authentication of updates.
However, $ID_{dst}$ should also be periodically changed (e.g. once every couple of
hours) and its locators updated in order to improve the robustness of the scheme, since this identifier
and locator will be included in clients' certificates.
The receiver will find the sender's identifier and locator in the first received
packet, and will only trust them, after executing the below explained checks.

% \commenti{Explicar que obter o ID do receptor é um problema do ovo e da galinha que tem várias soluções possíveis (DNS, tracker). Esta assimetria não é grave dd que seja possível obter o id da outra parte com confiança}

Since communication is now directed at identifiers instead of IP addresses,
there is now the matter of obtaining and resolving them to locators. For public-facing servers,
the DNS and the LISP Mapping system may be used as explained above.
In the other cases, the initial destination's
identifier and locators should be obtained
from trusted third parties (\eg by making use of a tracker in a P2P system).

The entities involved in communication have at their disposal one or more pairs
of public/private keys ($Pub_{\text{key}}$, $Priv_{\text{key}}$), used without
restrictions. The certification of such public keys by third parties is not mandatory,
being orthogonal to this proposal. It is assumed that the peers' clocks are
synchronized with Universal Time Coordinated (UTC), with an error smaller than
$Clock_{error}$.

When a client wishes to communicate with a server with $ID_{dst}$ and $LOC_{dst}$, it
generates a certificate $Cert_{src}$ and $ID_{src}$:

{\small

  $$ Cert_{src} = (LOC_{src}, LOC_{dst}, ID_{dst}, Duration_{ID_{src}}, time, ClientPub_{key}, Sig) $$
  $$ ID_{src} = LSB_{\#bits}(HMAC_{k}(Cert_{src})), \text{where } k = H(ClientPub_{key}) $$
}

$Sig$ is a digital signature based on $ClientPriv_{key}$ over the remaining
certificate parameters. $HMAC_k()$  is a Hash based Message Authentication Code using
the key $k$. $H()$ is a secure hash function. $LSB_{\#bits}()$ denotes the
least significant $\#bits$.
Later on we will present the selected
specific cryptographic functions used for implementation purposes.
The identifier $ID_{src}$ is generated from the context information in the
certificate, which means it is specific to that transaction, being valid for
$Duration_{ID_{src}} + 2 \times Clock_{error}$ since the moment \emph{time}. Nevertheless, the
receiver may refuse to accept the certificate if $Duration_{ID_{src}}$ exceeds a threshold value
chosen by himself.
% The maximum value of $Duration_{ID_{src}}$ and $Clock_{error}$ is set in a
% standard, and their value is assumed to be 10 and 2 minutes, respectively.

The client's first packet is sent from $ID_{src}$ to $ID_{dst}$ through the LISP tunnel
established between $LOC_{src}$ and $LOC_{dst}$. The certificate $Cert_{src}$ is transmitted
in the security header. Upon receiving a packet, the receiver makes the following checks:

\begin{enumerate}
\item $ID_{dst}$ belongs to its own set of identifiers
\item $Cert_{orig}$ is temporally valid (\ie{} hasn't expired and $time$ is not in
the future)
\item $ID_{src} = HMAC_{k}(Cert_{src})$
\item The signature $Sig$ in $Cert_{src}$ is valid
\end{enumerate}

If the security header is valid and the request is immediately
accepted,  the reply follows
in a packet from $ID_{dst}$ to $ID_{src}$ over the LISP tunnel established between
$LOC_{dst}$ and $LOC_{src}$.

% Impact on DDoS attacks/Why is this useful

As a result of the described mechanism, the victim of a DRDoS
attack is able to unambiguously identify the packets belonging to the
attack, at the cost of comparing the packets' identifiers with its own, allowing
effective filtering of attack packets. This identification and filtering may be
delegated onto trusted third parties (\eg providers edge devices).
%such as simple security equipment close to the victim.

However, the reflectors maintain a passive role, checking only the validity of
the new headers, while the victim is still subject to the volumetric aspect of
the attack. Moreover, packets must be processed, even if it simply means discarding
them. Ideally, the reflectors would have a more active role in mitigating
attacks.

% Description of the challenge mechanism

For this end, we optionally make use of receiver generated challenge puzzles or {\rgp}s.
{\rgp}s are cryptographic puzzles posed by the
receiver to the party who initiates a transaction. Upon receiving a packet from
an hitherto unknown party, the receiver replies with a {\rgp},
requiring its successful resolution before processing the initial request.
The introduction of this mechanism provides two new advantages.

Firstly, if a party truly wishes to have its request processed, it must
correctly solve the challenge posed by the remote party. To do so, the packets'
source location must be the real one, otherwise the challenge packet won't be received.
As a consequence, all peers must use their real locators, which makes it
impossible to perform reflected attacks. Although it is possible to use the
challenges as an attack vector, it would be both less effective than current
vectors, and easier to detect and mitigate.

Secondly, it allows the receiver to distinguish between hosts who have solved a
challenge (revealing their real location and proving some goodwill). In turn,
this allows the usage of more fine-grained traffic shaping rules (see Section~\ref{evaluation}), specifically
targeting hosts generating a high volume of unsolved {\rgp}s, which
is rather suspicious.
The effort asymmetry in posing vs. solving the puzzle may also provide some
degree of protection against SYN flood attacks, and application layer attacks.

% Description of the adopted puzzle

A {\rgp} solution requires a brute-force hash collision computation, inspired by
\cite{parno_portcullis:_2007} and \cite{nygren_tls_2015}.
Upon receiving a packet from a non-trusted host, the receiver generates a
pseudo-random number $n$ (which the sender must guess) with $64 + 2^{l}$ bits,
and sends the receiver a packet with:

\begin{itemize}
\item the length $l$ of the secret number $n$,
\item the partial solution $n'$, which is simply $n$ with the $K_{bm}$ least significant
bits masked,
\item the length $K_{bm}$ of the bit mask,
\item the hash $h = H_c(c + ID_{src} + ID_{dst})$, where ``$+$'' denotes concatenation
\end{itemize}

The client will have to guess, by brute-force, a candidate solution $c$ (of
which $n'$ is a prefix), such that $h = H_c(c + ID_{src} + ID_{dst})$.
When such a candidate solution is found, the sender
should retransmit the initial request to the receiver, as well as the
challenge and the candidate solution.
If the receiver is able to confirm the validity of the candidate solution, it then
processes the request as usual.

To reduce the viability of challenge harvesting, the following
precautions are adopted:

\begin{itemize}
\item Including $ID_{\text{dst}}$ and $ID_{\text{src}}$ in the hash reduces its
scope to those particular hosts while the identifiers are valid.
\item The secret number generator should use a frequently rotated, ephemeral private
key, making it unlikely that the same secret number would work on puzzles
posed by different hosts.
\end{itemize}

All the above tests can be skipped if the receiver so desires. On one hand,
it may be very lightly loaded. On the other hand, the reply may be shorter then the
request so as to lower any reflection peril.
Also, note that the receiver, after the initial handshake,
if it finds the source identifier, $ID_{src}$, of subsequent sender packets,
in its own trustable identifiers whitelist, it will also bypass all checks.
The validity of each whitelist entry is limited by the time-to-live
of the respective certificate.

Table \ref{tab:crypto} summarizes the cryptographic functions selected to
implement the {\eipi} proposal.
Note that the cryptographic functions selected offer at least the same
security guarantees than the current cryptographic suite
used by IPSecv3~(RFC6071) and IKEv2~(RFC7296) as we will discuss
in Section~\ref{costs}.

\begin{table}
 \centering
 \caption{Cryptographic functions selected for {\eipi} implementation}
 \label{tab:crypto}
 \footnotesize
 \begin{tabular}{c|l|l}
 \midrule
\textbf{Function} &\ \ \ \  \ \ \ \ \ \ \ \ \ \  \ \ \textbf{Usage} & \ \ \ \ \ \  \ \  \ \ \ \ \ \ \ \ \textbf{Cryptographic algorithms}\\
\midrule
$HMAC_k()$ & Identifier generation & HMAC~(RFC2104) using $SHA3_{256-121}$  \\
                      &                                  & \ie producing 256 bits truncated by using $LSB_{\#bits}$ \\
                       &                                 & with $\#bits = 121 $ for ORCHID compliance \\
%   & & \\
Sig & Certificate digital signature & RSA signature using PKCS\#1 with 1024 bits key \\
% & & \\
$H(), H()_c$ & Identifier and {\rgp} generation &  $ SHA3_{256} $ \\
\midrule
\end{tabular}
\end{table}

\section{Attack mitigation effectiveness}
\label{evaluation}

%%%%%%%%%%%%       \subsection*{Goals, methodology and model}

%As already stated, distributed reflected denial of service (DRDoS) attacks are the most common type of DoS attack. In the past, these attacks focused on exploiting vectors that would increase the attack's bandwidth (amplification attacks), although the most recent trends point to a growing reliance on a high packet rate \cite{akamai-q3-2015}. Thus the model will need to have the attack bandwidth and
%the amplification factor as parameters.

%%%% The protocol's effectiveness against application-level attacks will also be considered.

Obtaining attack traces that are simultaneously recent, publicly available, and representative of complete attacks isn't an easy task.
This is compounded by the fact that attackers may employ different approaches during the attack
(such as changing the attack vector, focusing on bandwidth vs. packet rate, and rotating the botnet's live population).
Thus, a simpler model based on the worst case scenario for the defender, \ie the most violent points in time of a DRDoS attack,
was used to assess {\eipi} effectiveness.
That model will have the attack bandwidth and
the amplification factor as parameters.

% \subsection{Model and results}

%%%%%%%%%%%%%%%%%% \commenti{rate or bandwidth ?}

The attacker can generate attack packets at the average bandwidth or rate $R_{a}$ bps,
sent by servers hosted by cloud providers, or by bots with a smaller upload capacity.
%A bot member upload bandwidth  can be estimated as a tenth of the global average peak download bandwidth,
%or around  a tenth of 30 Mbps as reported by Akamai in 2015 \cite{akamai-2015}.
%Thus, by that time, around 300 bots are required to generate 1 Gbps of attack bandwidth.
By hypothesis, without {\eipi} these packets will be accepted by reflectors as genuine.
With {\eipi} they include a certificate and successfully pass the identifier and certificate checks.
Thus, the attacker sends packets of dimension $D_{req}$ bits
in a traditional or baseline setting,  and packets including a certificate of dimension $D_{req+cert}$ bits
when using {\eipi}.
We assume (on average) small request packets such that $D_{req}= 100 \times 8 = 800$  bits including headers.
According to the cryptographic choices in Table~\ref{tab:crypto},
$D_{req+cert} = 450 \times 8 = 3600$ bits and the dimension of a puzzle challenge
is $D_{puz} = 272 \times 8 = 2176$ bits.
The size of the reflector answer, if any, is $D_{rep} = D_{req} \times A_{f}$
if the reflector directly answers the request with an amplification factor $A_{f}$, or
$D_{puz}$ if the reflector resorts to making use of {\rgp}s.

% In the sequel we will deduce the flooding bandwidth received by the victim $R_{v}$ function of $R_a$.
We will later deduce the flooding bandwidth received by the victim $R_{v}$ as a function of $R_a$.
Recall that the victim will easily and surely
identify the attack packets as bogus when using {\eipi}. The following scenarios will be explored:

\begin{enumerate}
\item A baseline scenario % representing the current state of the Internet,
with no packet filtering techniques;
\item The second scenario introduces the simplest form of the protocol, with identifier and signature validation, but no client puzzles. All packets sent by the attacker include valid security headers (otherwise the receiver or any other intermediate third party could discard the packets as invalid), and all receivers (reflectors) check the validity of both the identifier and the signature;
\item The third scenario builds on the second scenario: besides verifying the header validity, the reflectors
also resort to making use of {\rgp}s. Since this is a reflection attack,
the source locator in the packets sent by the bots is the victim's. As a result, the victim
will be flooded by reflectors with {\rgp} packets it did not request.
\item Finally, the fourth scenario builds on the third and introduces a shaper policy in every reflector. This policy limits
the traffic sent by each reflector, thus, the attack bandwidth is proportional to the number of reflectors.
$R$ is the number of reflectors and $R_{shap}$ the average rate of outstanding {\rgp}s
that each reflector accepts per locator bucket (\eg a /24 for IPv4
and a /56 for IPv6), the so called peak packet hit rate allowed by the shapers policy.
Lastly, $R_{v}$ is also computed with $R = 10^3$ and $R_{shap} =  10$, \ie
1000 reflectors and each one accepts at most 10 unanswered puzzles per  locator bucket
on average.
%\footnote{
The first parameter seems a good guess, the second is similar to the shaping
parameter used by the RRL (Response Rate Limit) \cite{dns-rrl}
mechanism of DNS servers. % to prevent DRDoS attacks making use of them.
% }.
\end{enumerate}

%%%%%%%%%%%%%   \subsection*{Results}

\begin{figure}[!htb]
    \centering
    \begin{minipage}[t]{.45\textwidth}
        \centering
        \footnotesize
\includegraphics[scale=0.45]{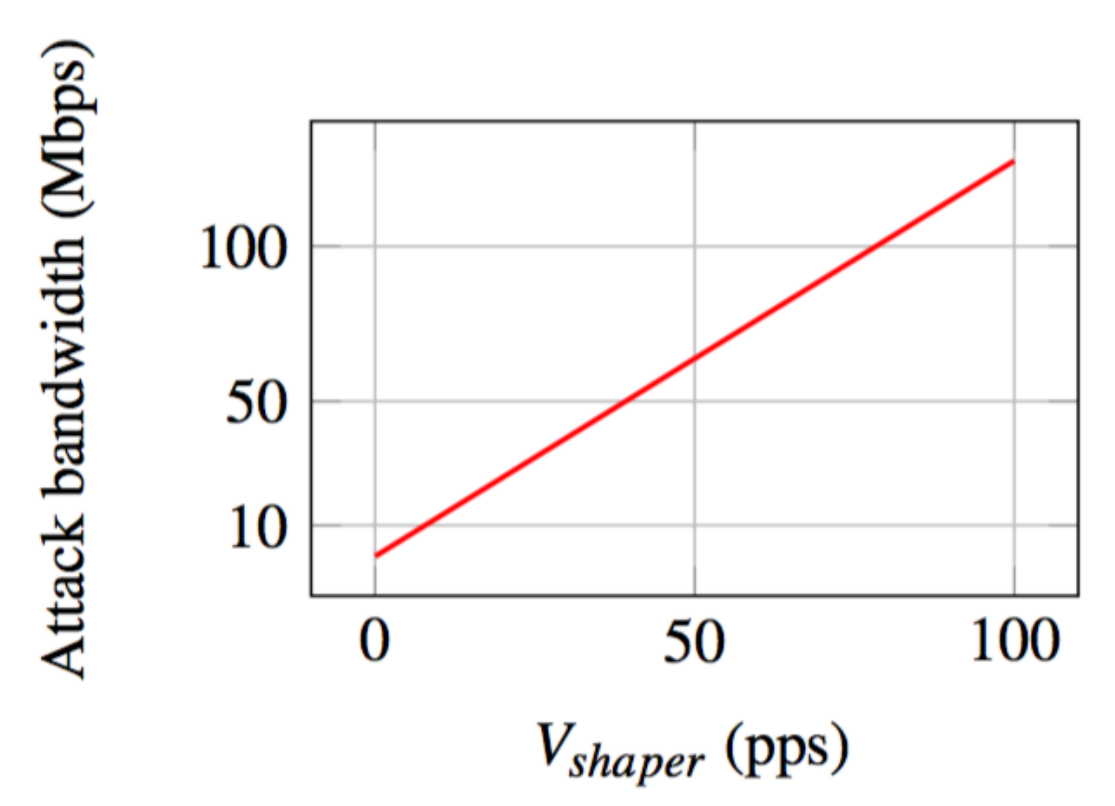}
        \caption{Attack bandwidth as a function of the peak packet hit rate allowed by the shapers policy per 1000 reflectors.}
        \label{fig:attack-shaper}
    \end{minipage}%
    \hspace{5mm}
    \begin{minipage}[t]{0.45\textwidth}
        \centering
         \footnotesize
\includegraphics[scale=0.45]{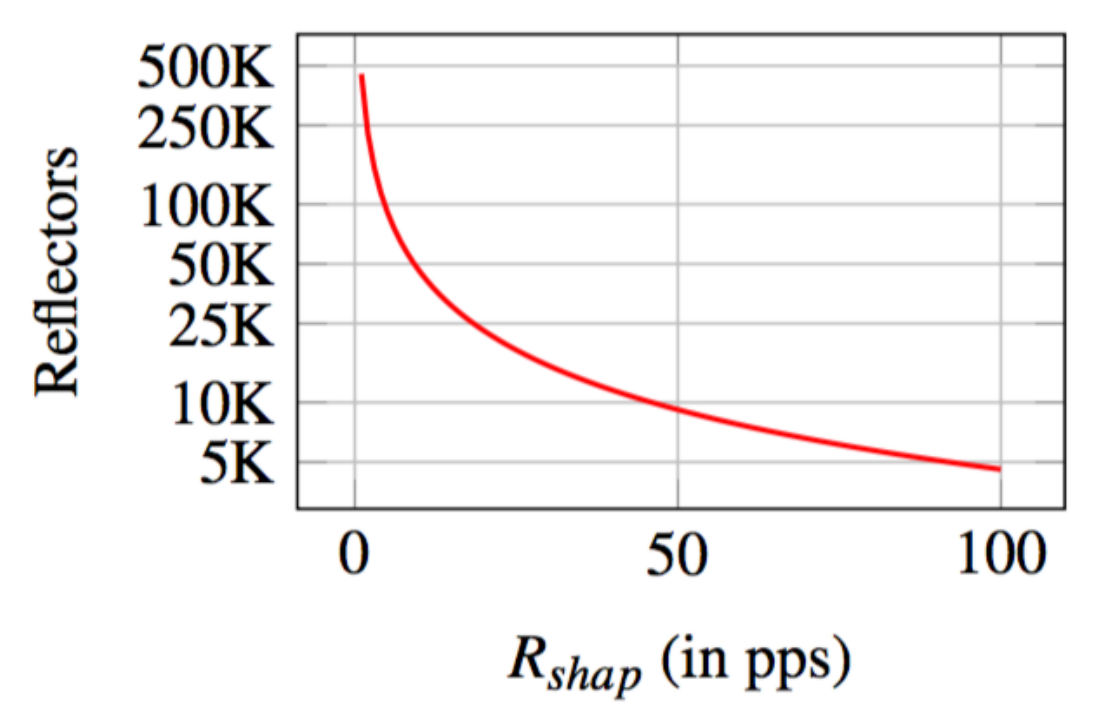}
         \caption{Reflectors required per Gbps of attack bandwidth, as a function of the peak packet hit rate allowed by the shapers policy.}
         \label{fig:reflectors-per-gbps}
    \end{minipage}
\end{figure}

The attack bandwidth $R_v$ as function of the attacker bandwidth $R_a$ is presented in table \ref{tab:performance-proposal} in the
different scenarios. The last column of the same table presents results when  $R_a = 1$ Gbps. In scenario 4, the results
are independent of $R_{a}$ and are only related
to the number of reflectors $R$ and the peak packet hit rate allowed by the shapers policy $R_{shap} / D_{puz}$.
Figure \ref{fig:attack-shaper} represents the attack traffic $R_v$ as a function of the packet hit
rate allowed by the shapers policy, when using 1000 reflectors.

\begin{table}
 \centering
 \caption{Attack bandwidth $R_{v}$ generated by an attack of rate $R_{a}$.}
 \label{tab:performance-proposal}
 \footnotesize
 %\tiny
 \begin{tabular}{l|c|c|c|c}
 \textbf{Scenario} & \ \ \textbf{Reflectors} \ \ & $ \ \  \mathbf{R_{shap}}$\ \  & $ \ \ \mathbf{R_{v}}$ & $ \ \ \mathbf{R_{v}} \ \  (R_a = 1$ Gbps) \\
%                           &                                &      		                          &   	  &   \\
\midrule
Baseline           & 	\--- &		\---	         & $ R_a / D_{req} \times D_{req} \times A_{f} $ 	& $1 \ $ Gbps $ \times A_{f} $  \\
2                       &      \--- &		\---	         & $ R_a / D_{req+cert} \times D_{req} \times A_{f}$	& $ 222 \ $ Mbps $\times A_{f} $  \\
3                       &      \--- &		\---	         & $ R_a / D_{req+cert} \times D_{puz} 		      $ & $  604 \ $ Mbps    \\
% 4                  	&   $R $ 		&  $ R_{shap} 	       $ & $  R \times R_{shap} \times D_{puz}		      $	 &           \\
% 			&			&				  &										 & \\
% 4                  	&   1000 &    10  & $  					      $	 & 12.76 Mbps         \\
4                  	&   1000 		&  10 	        & $  R \times R_{shap} \times D_{puz}		      $	 & 12.76 Mbps          \\
\end{tabular}
\end{table}

Figure \ref{fig:reflectors-per-gbps} represents the cost to the attacker of 1 Gbps of attack traffic
in scenario 4, expressed in the number of reflectors required to perform the attack as a function
of the packet hit ratio of the reflectors shapers. For instance, an attack of 12 Gbps
(described as average by several sources \cite{akamai-q3-2015,verisign2014}) would require about
1 million reflectors with $ R_{shap} = 10$ packets per second (pps).

%%%%%%%% MINIPAGE WITH BOTH GRAPHS %%%%%%%%%%%

%%%%%%%%%%%%%%%%%%  \subsection*{Discussion and conclusions}

When the protocol is used, both the locator and the identifier of a remote peer are needed to exchange packets with it, since a host rejects packets addressed to an incorrect (even if valid) identifier. As a result, an attacker would be unable to obtain new reflectors to perform a DRDoS attack by simply scanning a range of IP addresses, and would have to rely more on hosts with public, stable identifiers. With low rate shapers, a
successful attack will require a huge number of reflectors, unlikely to be gathered in due time, since their identifiers are valid only during a couple of hours.

%%%%%%%%%%%%%%% \subsection{Transport and application-layer attacks}

The protocol is able to effectively mitigate SYN floods when client puzzles are used, which can be viewed as a generalization of the SYN Cookie mechanism.
Additionally, the client puzzles may also mitigate application-layer DoS attacks, by imposing increasingly harder client puzzles to misbehaving peers. However, some precautions must be taken against indirect attacks with the purpose of raising the difficulty of client puzzles for innocent third parties.

%%%%%%%%%%%%%%    \subsection{Progressive adoption}

The protocol may be progressively adopted since servers may continue to accept requests directly sent to traditional
IP addresses during a transition period. Additionally, they can also accept requests without security headers sent to their identifiers by the way of tunnels.
In both situations, if servers shape the amount of requests with non verifiable source address integrity, they may increase the
incentive for the adoption of {\eipi}.

% \begin{figure}[htpb]
%   \centering
%   \footnotesize
%   \begin{tikzpicture}
%     \begin{semilogyaxis}[
%       width=0.45\textwidth,
%       height=4cm,
%       xlabel={$R_{shap}$ (in pps)},
%       ylabel={Reflectors},
%       ytick={5e3,1e4,2.5e4,5e4,1e5,2.5e5,5e5,7.5e5},
%       yticklabels={5K,10K,25K,50K,100K,250K,500K,750K},
%       grid=major,
%       ]
%       \addplot[domain=0:100, samples=100, thick,red] {10^9/(x*272*8)};
%     \end{semilogyaxis}
%   \end{tikzpicture}
%   \caption{Reflectors required per Gbps of attack bandwidth, as a function of the peak packet hit rate allowed by the shapers policy.}
%   \label{fig:reflectors-per-gbps}
% \end{figure}

%%%%%%%%%  \subsection{Conclusions}

% \commenti{Chamar a atenção que o protocolo é independente do vector de ataque, requer uma qtd maior de reflectores, dificulta a colecta de reflectores, pode ser delegado em terceiros, e tem repercussões nos diferentes níveis da pilha}

The protocol works at the network level, making it independent of the attack vector\footnote{A few exceptions remain, such as the so-called ``quiet attacks'' \cite{DBLP:journals/icl/ShevtekarA09}, directed at routers.}.
It introduces a new requirement to successfully exchange packets with other hosts, making it harder to harvest commonly available reflectors (\eg{} home routers, \ldots), while simultaneously requiring more reflectors to perform a similarly-sized attack.
Candidate reflectors do not need to always resort to making use of challenge puzzles. For example,
when the protocol at hand is UDP and the reply implies $ A_f < 1$ the potential resulting attack bandwidth is negligible.

The handling of the protocol may also be delegated into trusted third parties, thereby negating the computational costs to servers. Finally, the client puzzle may be seamlessly integrated with the TCP handshake, removing the need for additional packet exchange.

%\commenti{Haverá mais consequências noutros níveis da pilha?}

The shaper policy plays a large role in the protocol's effectiveness. However, even a lax policy like 50 client unanswered outstanding puzzles per second per source locator bucket, when using 1000 reflectors, brings the attack bandwidth down from 1 Gbps $\times A_f$ to 63.8 Mbps, and the attack packet rate down from 12.5 Mpps to 0.05 Mpps  --- a reduction of nearly 93.6\%  (in the absence of amplification!) and 99.6\%, respectively.

\section{{\eipi} security and computational costs}
\label{costs}

\textbf{Security considerations}. The {\eipi}  cryptographic suite was selected to offer at 
least the same security guarantees as provided by the last IPSec related standardization (RFCs 6071, 4302 and 4303). 

RFC 6071 presents the IPSecv3 and IKEv2 roadmap and cryptographic updates in other specific RFCs. HMAC-SHA1-96 (for hash-based macs) and AES-XCBC-MAC (using AES cryptographic MACs) are standardized alternatives for ESP message integrity verifications (in the RFC 4308), for two configuration settings for Virtual Private Networks. These suites are intended to be used in IPSec as single-button choices for alternative VPN configuration purposes. The RFC 4869 (defined by NSA) recommends the use of AES-GMAC and HMAC-SHA functions for ESP integrity control.  
The RFC 6379 updated the RFC 4869 promoting HMAC-SHA-256-128, and HMAC-SHA-384-192, specifically for integrity checks in IKEv2 (for IPSec Key-Exchange), as previously recommended in the RFC 4869 US NSA proposal. 
Nevertheless, it must be noticeable that those proposals refer HMACs with hash functions from the SHA-2 family. 

In the {\eipi} proposal we decided to evaluate the computational costs of integrity checks with HMAC-SHA3-256 (using the SHA-3 family or Keccak family)~\cite{nist2015}, comparing with an optional use of AES-GMAC-256 (cryptographic MAC using AES and a 256 bit key). Although this is not necessarily a notorious security improvement comparing to the SHA-2 family (as used in IPSec), our evaluation is aligned with the recent enforcement from NIST and FIPS-PUB standardization in promoting the recently standardized SHA-3 Hash-Functions and variants. 

\textbf{Experimental evaluations.} 
For our experimental evaluations we used HMAC-SHA3-256 and AES-GMAC-256 cryptosuites implemented in the Openssl library v0.98za, running in a Intel Core i7 \  2.5GHz QuadCore \  CPU and MAC-OS X Mavericks v10.9.5. In this testbench, RSA and DSA integer exponentiation and multiplication are performed by making use of hybrid software and hardware acceleration co-processing. All running time
evaluations were performed by repeating each computation 1,200 times,
 in 12 series of 100 trials, performed at different moments. To avoid
 most noise, during tests, other computational tasks in the 
 same computer were avoided. Results series exhibit a standard deviation
 of at most 1\%.

A candidate attacker, as well as a regular user, will need 470 ms to generate
a certificate (using an already available RSA 1024 bits key pair since these can
be prepared in advance). The defender only needs 37 ms to verify the 
certificate signature. Recall that the attacker needs $ID_{dst}$ to 
produce the certificate and this is only valid for a couple of hours. Making use
of the HMAC function to compute $ID_{src}$ requires 13.98 $\mu$s,
and 13.81 $\mu $s to verify, almost the same time.

In our practical observations we repeated all the above tests using AES-GMAC-256
(using a 256 bit key), as a possible alternative for HMAC-SHA3-256, exactly the same
integrity proof standardized for IPSec in RFC 6071 and ESP in RFC-6379.
The impact of using this cryptosuite is to slow (in average by a factor of 2,72) the computational costs, comparing with HMAC-SHA3-256. Then, in the same experimental settings the computation of $ID_{src}$ requires 53,26 $\mu$s and 52,06 $\mu$s to verify, a balanced time for generation and verification. This includes the key generation from an initial seed and the CMAC computation itself.

\textbf{Resistance against $ID_{src}$ forging.} To be successful in a reflection
attack with violation of the integrity of the
identifiers, an attacker should be able to produce a collision with $ID_{src}$ 
(which is only valid during a couple of hours) using a ``valid certificate''. 
Again, the best scenario for the attacker is when the receiver only
verifies that $time$ is valid and $ID_{src} = HMAC_{k}(Cert_{src})$
without verifying the signature.

For this to happen, the attacker needs to violate in the best effort the property of 
\emph{second-image resistance} of the SHA-3 function implicit to the HMAC computation. 
Since there are no publicly known cryptographic vulnerabilities of the SHA-3 function ``sponge'' and
the synthesis primitive Keccak embedded in SHA3 \cite{sha3} the attacker needs to generate
$2^{128}$  ($\approx 3,4 \times 10^{38}$) identifiers,
requiring $1.32\times10^{30}$ hours in a single processor like the one used in tests,
and $1.32\times10^{10}$ hours (around one million years) when using $10^{20}$ (1000 billion)
computers, to perform an operation whose usefulness lasts for a couple of hours.
One may note that only the least 121 bits of the HMAC are required, and not all
the 256 bits of the output. Any way, the probability of occurring a collision of the
least significant 121 in 256 bits pseudo randomly generated is $\approx 10^{-32}$. This value 
may be computed using formula $1-e^{-(N\times(N-1))/2^K}$, with N trials and K=121 (bits)
according to appendix 1 of~\cite{stallings_computer_2014}.

By making use of  AES-GMAC-256, the attacker must invest in 
generating $2^{256}$ different identifiers (by brute-forcing all possible AES keys generated from the unknown seed), or $9.95*10^{70}$ 
hours using a single processor or around $9,95*10^{50}$ hours using 1000 billion computers. 
The other way is to compute the identifier using some form of cryptanalysis against AES (by breaking the AES computation and the 256 bit key) by plaintext/cyphertext combinations, a not feasible attack in the time to live of valid $ID_{dst}$ identifiers.

\textbf{{\rgp} solving.} In order to solve the {\rgp} sent by the receiver in the challenge message, the
initial sender of the first packet needs to find by brute force
the integer $c$ from which it already knows 
all bits except $K_{bm}$, the length of the bit mask of the puzzle, 
such that $H = SHA3_{256}(c+ID_{orig}+ID_{dst})$
is the result of the challenge. 

In the same hardware, the verification of the solution
by the challenger is the cost of a comparison if it stores all the challenges sent
recently, or the cost of computing an HMAC if not, \ie 13.81 $\mu $s to verify.
The solution of the puzzle requires on average $2^{K_{bm}/2}$ trials according to the
birthday collision paradox. In the same setting, with $K_{bm}=20$ it takes 
approximately 13 ms.

For the case of AES-GMAC-256, the sender of the first packet needs now to find $c$ from which it already
knows all bits except $K_{bm}$, the length of the puzzle, such that 
$H = AES-GMAC-256(c+ID_{orig} + ID_{dst})$ is the correct result for the challenge. 
The  verification of the solution by the challenger is the comparison cost of 
storing all the challenges sent recently, or the cost of computing on-the-fly the AES-GMAC-256, \ie only 51.51 $\mu$s to verify. 
Comparing with the HMAC-SHA3-256 option, the cost is a little bit higher for the defender, but even more unbalanced in the effort required to the attacker.

In conclusion, we emphasize that our experimental observations above are the best choices reflecting the best case scenario for attackers intending to send incorrect identifiers, with no significant cost for an honest receiver.

\section{Related work}
\label{related}

DoS combat effectiveness can not be tackled by
 ``black-and-white'' arguments.
On one hand, several possible measures are not realistic since they
increase the discomfort of users and quickly become incentives to circumvent them,
what turns upside down their initial goal. On the other hand, approaches
must be balanced against investments and operational costs increase,
complexity increment,
as well as architectural decisions leading to diminishing scalability returns.
Moreover, and above all,
incentives play a pivotal role in DoS combat assessment. Many proposals increase
costs where there are no incentives to implement them and are therefore
doomed to fail.

Defensive walls against DoS can be implemented near the attackers and reflectors,
near the victims or in the core of the Internet \cite{survey-ddos1}. One of the
easiest ways to lower DoS impact would be to prevent ISP
(Internet Service Providers) customers
from making use of IP source address spoofing, as is recommended in
BCP 38 (``Best Current Practices") \cite{rfc2827} since 2000. However,
by lack of incentives, this practice is today far from being generalized.
%Thus, proposals that require the setup of complex  systems near the
%source of the attacks (\eg reverse firewalls, distributed alarm systems, \dots)
%are not very realistic.

Easier to implement are measures
preventing critical public infrastructure servers from being used
as reflectors. For example, most publicly available DNS servers
make use of request dampening  and filtering
(\eg Response Rate Limit - RRL \cite{dns-rrl}) techniques to
prevent their usage as DDoS reflectors. However, the available
methods are all protocol dependent and are based on heuristics inspired
from the current attack practices, which may be circumvented by
more sophisticated attackers \cite{report-rrl}.
Proposals that increase the complexity and costs of the core
(\cite{survey-ddos1}) are also not very realistic since the core is where scalability plays
a paramount role. Moreover, transit operators lack the incentives to implement them.

This leaves defensive walls near the victims as the most realistic.
These walls are today of two types: firstly, capacity and diversity increase
and, secondly, inbound traffic filtering or shaping.
Capacity increase is an expensive and
never ending type of measure whose effectiveness against amplification
attacks is arguable. Diversity can be implemented by making use of cloud
or CDN (Content Distributed Networks) providers that require
attackers to spread their forces and targets. Additionally, many of these providers
offer DoS oriented filtering and forensic services invoiced as an extra.
Ironically, it has been observed that most publicly contractable booter
services are protected from the attacks of other booters by these types
of services \cite{booters}.

%Filtering and dampening techniques are implemented by (mostly) application-level firewalls
%that track inbound traffic for the presence of attack signatures \cite{survey-ddos1},
%and block or shape flagged packets.
%Detecting all protocol attack vectors is a long term goal
%requiring massive (deep) packet inspection, history analysis, and
%sophisticated engineering muscle, which often resorts to machine learning
%methods. The performance of these methods is inverse to the number
%of false negatives that cross the wall, as well as to the number of
%false positives they block, which prevent honest users from using the
%service, thus also increasing DoS attacks success.
%
%DoS prevention can be implemented at several layers: application, transport or network.
%For example, IP source address spoofing prevention is a network layer measure.
%Response rate limiting as well as most filtering techniques are application layer measures.
%The lower the layer used to detect the attack, the
%most general purpose and attack vector independence the method is,
%thus increasing its effectiveness.
%

%{\eipi} is implemented at the network level, is transport and application level protocols independent, can be
%progressively adopted, tries to put the incentives at the right stake-holders
%(public-facing services, potential bots and reflectors), pushes clients for
%its adoption since they can be better (or only) served if they make use of it, and is not
%dependent of third parties. We present below an analysis of similar proposals
%using the above analysis framework.}

We now turn to the more directly related work analysis.
Mandatory IPSec adoption would make users dependent of third parties
for certificate authentication and would increase user costs.
Moreover,
it would increase the potential for security guaranties concentration in a,
necesserely small, number of certificate issuing operators responsible for security
at all layers: application, transport (\eg TLS) and application.
Finally,
it would increase behind acceptable limits the latency of all short-term interactions.
Our proposal is also a network layer one. However, by relying on a
locator / identifier separation proposal ({\mapenc}) using ephemeral self-certified
identifiers and ephemeral self-certified association between identifiers
and locations, it aims at avoiding the complexities of certificate management and
overhead increase issues.

All {\mapenc} proposals implement the authenticity of the association between
(prefix) identifiers and their locators. LISP, as well as other
proposals (\eg \cite{ddos-ssm}), delegate this pivotal security role to the mapping
system, which becomes a central security component. This centralization is
compatible with today's AS and ISP ecosystem, but may become a
nightmare against competition and privacy in a Internet dominated by
mobility and roaming individuals, receiving simultaneous service from several different providers
(\eg cellular, WiFi, \dots). Moreover,
all locators updates can only take place by using the mapping system as
a mediating party, which is an additional factor of latency increase during mobility events,
or the activation of new interfaces.

With {\eipi}, only public servers are required to have their locators
published in the mapping system, since it introduces a new end-to-end mechanism supporting the
verification of the authenticity of the relationship between identifiers and
locators. For example, after the initial verifications and challenges,
both communication parties can use a lighter security protocol to update their locations,
without depending of third parties. In situations of low ``criticality level'',
the answer to an isolated query can be directly sent to the locator that
originated it.

Paper \cite{ddos-ssm} also proposes a {\mapenc}-based solution and claims
its effectiveness in DDoS prevention and detection.
However, this solution leverages the mapping system for all
packet exchanges among hosts, is restricted to tunnels ending
in network operators equipment,
is not end-to-end, restricts a host to operators that can authenticate it,
and only works if both
end-tunnel operators can authenticate each other.

Response Rate Limiting (RRL)~\cite{dns-rrl} is an application layer approach
specific to the DNS protocol.
Its effectiveness is also dependent of the way DNS clients currently perform queries
and would fail if attackers become more sophisticated~\cite{report-rrl}.

Host Identity Protocol (HIP) \cite{hip} is an end-to-end protocol
that makes use of hashed public keys as host identifiers. By using IPSec,
HIP allows two mobile hosts to establish a
secure and authenticated tunnel among them. It is heavier and requires
more packet exchanges  than our proposal since it achieves a higher security level
similar to IPSec tunnels or TLS connections. To combat DoS, HIP also uses puzzles during
the handshake.

Our proposal goes in the same vein as weak authentication \cite{arkko_weak_2004} and
besides the installation of a new software module in the system
kernel, it requires no further action from the user, since it does not aim
at solving all the security problems, but only introduces a light layer
of identifier and location integrity, without precluding the adoption of
any other complementary and higher level security measures.

Additionally, this small layer of address integrity and locality ascertainably and verification
can be delegated by public-facing servers in third party front ends
(\eg cloud operators), without
reducing server security or requiring the delegation of their identity
and authentication keys or certificates. In this environment, only clients would be forced to
directly support an increased overhead. This contributes to put the incentives in the
right place.

\section{Conclusions and future work}
\label{conclusions}

Nowadays DoS attacks represent a significant fraction of all
attacks that take place in the Internet, leading to significant
economic losses by comparison with attackers small investments.

Due to its root causes, this state of affairs can not be changed overnight.
Currently, most successful combat measures
include servers replication and servers bandwidth increase, as well
as resorting to security providers that use expensive
and attack-vector dependent detection techniques in a never ending
``cops and robbers'' chase.

In this article we propose {\eipi}, an approach based on the use of
forge-resistant, self-certified, ephemeral identifiers as source addresses, in order to
allow a victim to easily discern packets from a reflected attack. Those
identifiers are optionally complemented with a client puzzle mechanism that also confirms a
client's location before processing their request.
The protocol involves an additional security header added to the
recently proposed  {\mapenc} approaches (\eg LISP),
containing context metadata. Thus, it can be
incrementally adopted, and requires no changes to either applications or
transport protocols. One main characteristic of the proposal is that it allows a fine adjustment of
the level of precautions taken by the defenders, thus introducing adaptability to the
context, a feature often absent in other security mechanisms.
Finally, it works at the network level, making it independent of the  attack vector.

{\eipi} introduces a new requirement to successfully exchange packets with other hosts,
making it harder to harvest commonly available reflectors (\eg{} home routers, \ldots),
while simultaneously requiring more reflectors to perform a similarly-sized attack.
Candidate reflectors do not need to always resort to making use of challenge puzzles. For example,
when the protocol at hand is UDP and the reply implies a contraction
of the query volume, the potential resulting attack bandwidth is negligible.
However, if the communication parties need to engage in a longer
interaction (\eg a TCP connection), {\eipi}
the TCP handshake, preventing SYN Flood attacks and making use
of {\rgp}s to adjust the client's computational effort to the current server context.

The proposal motto is to perceive security measures as a continuous
set of mechanisms and policies, from cheap and simple, to expensive and complex,
so as to balance risks, costs and incentives to attain real effectiveness. Moreover,
this proposal does not precludes the possible complementary use of other
mechanisms, namely those at the transport and application layers.

The simple worst-case analysis presented shows that the proposed
mechanisms are able to drastically reduce the power of the most
deadly popular attacks, thus fulfilling the claimed properties.
Future work includes an evaluation and refinement of {\eipi} based on a more multifaceted
and deeper analysis. It will also embrace a comprehensive study of {\eipi} implementation issues
and adoption incentives.

\tiny

\bibliographystyle{abbrv}

\bibliography{bibliography,rfc}

\end{document}